\documentclass[rnote]{aa}

\usepackage{natbib}
\usepackage{psfrag, color, epsfig, rotating, pstricks, pst-node}
\usepackage{dsfont}
\usepackage{txfonts}

\bibpunct{(}{)}{;}{a}{}{,}

\newcommand{\average}[1]{\left\langle #1 \right\rangle}

\newcommand{\vt}{\vartheta}

\newcommand{\dd}{{\mathrm d}}

\newcommand{\ontop}[2]{
  \renewcommand{\arraystretch}{0.2}
  \begin{array}{c}
  #1 \\ #2
  \end{array}
  \renewcommand{\arraystretch}{1.0}
}
\newcommand{\lsim}{\ontop{<}{\sim}}

\newcommand{\Map}{{M_{\rm ap}}}
\newcommand{\Mapsq}{{M_{\rm ap}^2}}

\newcommand{\Omegam}{\Omega_{\rm m}}

\newcommand{\ee}{{\rm e}}

\newcommand{\J}{{\rm J}}
\newcommand{\HH}{{\rm H}}

\newcommand{\tmin}{\theta_{\rm min}}
\newcommand{\tmax}{\theta_{\rm max}}

\begin{document}

\title{E- and B-mode mixing from incomplete knowledge of the shear correlation}

\author{M.~Kilbinger \and P.~Schneider \and T.~Eifler}

\institute{Argelander-Institut f\"ur Astronomie%
  \thanks{Founded by merging
    of the Sternwarte, Radio\-astro\-nomisches Institut, and Institut f\"ur
    Astrophysik und Extraterrestrische Forschung der Universit\"at
    Bonn},
  Universit\"at Bonn, Auf dem H\"ugel 71,
  D-53121 Bonn, Germany}

\offprints{Martin Kilbinger, \email{kilbinge@astro.uni-bonn.de}}

\titlerunning{E- and B-mode mixing}
\authorrunning{M.~Kilbinger, P.~Schneider \& T.Eifler}

\date{Received / Accepted}

\abstract%
%
{%
}%
{%
  We quantify the mixing of the measured cosmic-shear E- and B-modes
  caused by the lack of shear-correlation measurements on small and large
  scales, arising from a lack of close projected galaxy pairs
  and the finite field size, respectively.
}%
{%
  We calculate the aperture-mass statistics $\langle M_{\rm ap,
    \perp}^2 \rangle$ and the E-/B-mode shear-correlation functions
  $\xi_{\rm E, B \pm}$ where small- and large-scale cutoffs are taken into
  account. We assess the deviation of the obtained E-mode to the true
    E-mode and the introduction of a spurious B-mode.
}%
{%
  The measured aperture-mass dispersion is underestimated by more than 10\%
  on scales smaller than 12 times the lower cutoff.
  For a precise measurement of the E- and B-modes at the percent level
  using a combination of $\xi_{\rm E, B +}$ and $\xi_{\rm E, B -}$, a
  field as large as 7 (2.4) degrees is necessary for ground-based
  (space-based) observations.
}%
{%
}%


\maketitle

\section{Introduction}
\label{sec:intro}

The observation of the correlation between the shapes and orientation
of high-redshift galaxies has become a well-established method of
studying the dark matter distribution on very large scales. Using
theoretical predictions for the power spectrum of the projected cosmic
density field, cosmological parameters can be obtained from the
measurement of cosmic shear, most notably the matter density $\Omegam$
and the amplitude of density fluctuations $\sigma_8$. Very recent
results include ground-based \citep{CFHTLSwide, 2005MNRAS.359.1277M,
  CFHTLSdeep, JBBD06} and space-based \citep{2004ApJ...605...29R,
  2005MNRAS.361..160H} observations.

Presently, cosmic-shear surveys are still limited by systematic errors
arising from the imperfect shape measurement of faint galaxies and a
deficient PSF correction. Extensive studies and comparisons between
different data analysis methods are being made \citep{STEP1} to find
sources of any contamination of the cosmological signal. A common
means to check for systematics in the data is the decomposition of the
measured shear field (or power spectrum) into the gradient- and
curl-parts (E- and B-modes), see \citet{2002ApJ...568...20C},
hereafter C02, and \citet{2002A&A...389..729S}, hereafter S02. Since
the image distortions caused by gravitational lensing are (to first
order) curl-free, the presence of a B-mode is a distinct imprint of
residuals not completely removed from the measurement. A B-mode of
cosmological origin can be caused by the intrinsic correlation of 
galaxy orientations \citep[e.g.][]{2001ApJ...559..552C,
  2002MNRAS.335L..89J}.  However, this effect is thought to be very
small for reasonably deep surveys \citep[e.g.][]{2004MNRAS.353..529H},
and it can be removed from the shear signal using photometric
redshifts or by modeling the intrinsic alignment signal that is
distinct from the shear signal \citep{HH03, KS02, KS03,
  2005A&A...441...47K}. A further non-negligible source of confusion
could be the recently discovered intrinsic shape--shear correlation
\citep{2004PhRvD..70f3526H}. This contamination also creates a B-mode
\citep{Heymans06}, and it is not yet clear how to correct for this
effect.

As we present in this paper, up to now the E- and B-mode decomposition
cannot be performed directly from the shear correlation as measured
from the data, but it involves an integral over the shear-correlation
function $\xi_\pm$ to either arbitrary small or infinitely large
angular separations.  However, the scales on which $\xi_\pm$ can be
measured are limited. On arc-second scales, the blending of closely
projected galaxy pairs prohibits a reliable determination of the shape
of those galaxies. On large angular scales, the measurement is
restricted by the finite field-of-view. These limits cause a mixing
of the E- and B-modes with the currently-used estimators, preventing
their clear-cut separation. Although this is a well-known fact, to our
knowledge this mixing has not or only inadequately been taken into
account to date.

In this paper, we quantify the mixing of the E- and B-mode due to the
small- and large-scale limits of the shear-correlation measurements.
In Sect.~\ref{sec:map}, the aperture-mass dispersion and in
Sect.~\ref{sec:xiEB}, the E-/B shear-correlation functions are
considered as measures of the E- and B-mode, respectively. Section
\ref{sec:cosm} briefly discusses the dependence on cosmology.
In Sect.~\ref{sec:summary}, we give a summary and offer
ways of clearly determining the E- and the B-modes.

\section{Aperture-mass dispersion}
\label{sec:map}

The aperture-mass statistics \citep{KSFW94,S96} is a smoothed function
of the convergence field,
\begin{equation}
  \Map(\theta) = \int \dd^2 \vt U_\theta(|\vec \vt|) \kappa(\vec \vt),
  \label{map}
\end{equation}
where $U_\theta$ is a compensated filter function, i.e.\ $\int \dd \vt
\, \vt \, U_\theta(\vt) = 0$. Equation (\ref{map}) can be written as an
integral over the tangential shear $\gamma_{\rm t}$, weighted with the
function $Q_\theta(\vt) = 2/\vt^2 \int_0^\vt \dd \vt^\prime \,
\vt^\prime \, U_\theta(\vt^\prime) - U_\theta(\vt).$ Moreover,
$M_\perp$ is defined as the weighted integral over the
cross-component of shear $\gamma_\times$,
\begin{equation}
  M_{\rm ap, \perp}(\theta) = \int \dd^2 \vt Q_\theta(|\vec \vt|) \gamma_{\rm
  t, \times}(\vec \vt).
  \label{mapgamma}
\end{equation}
%

The dispersion of (\ref{map}) \citep{1998MNRAS.296..873S} can be
written in terms of the convergence power spectrum $P_\kappa$,
\begin{equation}
  \langle M_{\rm ap}^2 \rangle (\theta) = \frac 1 {2\pi} \int \dd
  \ell \, \ell \, P_\kappa(\ell) \hat U^2(\theta \ell),
\end{equation}
where $\hat U(\eta)$ is the Fourier transform of the filter function
$u(\vt/\theta) = \theta^2 U_\theta(\vt)$.  The dispersion of $M_\perp$
is non-zero only if a curl-part or B-mode is present in the shear
field.  This is not the case for a purely gravitational shear signal
(to first order). Therefore, the measurement of $\langle M^2_\perp
\rangle$ is a test for systematic errors and/or intrinsic galaxy
alignment.

Two sets of filter functions have been extensively used for cosmic-shear
measurements and studies. One set are polynomial functions with finite support
\citep{1998MNRAS.296..873S},
\begin{eqnarray}
  U_\theta(\vt) & = & \frac 9 {\pi \theta^2}
  \left(1-\frac{\vt^2}{\theta^2}\right)\left(\frac 1 3
    - \frac{\vt^2}{\theta^2} \right)
  \HH\left(1-\frac{\vt}{\theta}\right); \;\; 
  \hat U(\eta) = \frac{24 \J_4(\eta)}{\eta^2};
  \nonumber \\
  Q_\theta(\vt) & = & \frac 6 {\pi \theta^2} \frac{\vt^2}{\theta^2}
  \left(1-\frac{\vt^2}{\theta^2}\right)
  \HH\left(1-\frac{\vt}{\theta}\right);
  \label{UQ-poly}
\end{eqnarray}
where $\HH$ is the Heaviside step function. The other set is related
to Gaussians (C02),
\begin{eqnarray}
  U_\theta(\vt) & = & \frac 1 {2\pi\theta^2} \left( 1 - \frac{\vt^2} {2\theta^2}
  \right) \ee^{- {\vt^2}/{2\theta^2}} ; \;\;\;\; \hat U(\eta) =
  \frac{\eta^2}{2} \ee^{{-\eta^2}/{2}} ;
  \nonumber \\
  Q_\theta(\vt) & = & \frac{1}{4\pi\theta^2} \frac{\vt^2}{\theta^2}
  \ee^{- {\vt^2}/{2\theta^2}}.
\label{UQ-Gauss}
\end{eqnarray}

The aperture-mass dispersion can be obtained by integrating over the
two-point correlation function (2PCF) of shear, $\xi_\pm$
(C02, S02),
\begin{equation}
  \average{M_{{\rm ap}, \perp}^2(\theta)} = \frac{1}{2}
  \int_0^\infty \frac{\dd \vt \, \vt}{\theta^2} \left[
    \xi_+(\vt) T_+\left( \frac{\vt}{\theta} \right) \pm 
    \xi_-(\vt) T_-\left( \frac{\vt}{\theta} \right) \right].
  \label{Map-xi-EB}
\end{equation}
The 2PCF are defined in terms of the tangential and cross-component of
shear, $\gamma_{\rm t}$ and $\gamma_\times$, respectively, where these
two components are measured with respect to the direction connecting
two points. The 2PCF can be written as functions of the power spectrum
as follows,
\begin{equation}
\xi_\pm(\vt) = \average{\gamma_{\rm t} \gamma_{\rm t}}(\vt) \pm
\average{\gamma_\times \gamma_\times}(\vt) = \int_0^\infty \dd \ell \,
\ell \, P_\kappa(\ell) \, \J_{0,4}(\vt \ell).
\end{equation}
The functions $T_\pm$ in (\ref{Map-xi-EB}) are given by
\begin{equation}
    T_\pm(x) = \frac 1 {2\pi} \int_0^\infty \dd t \, t \, \J_{0,4}(x
    t) \hat U^2(t);
  \label{T_pm}
\end{equation}
closed-formula expressions in the case of the above defined filter
functions can be found in S02 and C02. Expression
(\ref{Map-xi-EB}) is very useful since direct determination of
the aperture-mass dispersion from data is very inefficient due to
boundary effects and masked regions in the images. The 2PCF, on the
other hand, can be obtained easily from any survey topology. Note that,
in the case of the polynomial filter, the integral in (\ref{Map-xi-EB})
only extends to $2\theta$, since then $T_\pm(x) = 0$ for $x>2$.

Projected close galaxy pairs in the data have to be excluded from the
shear measurements. The shape of close pairs cannot be estimated
reliably because of the blending of the individual galaxy images. That
means that the 2PCF cannot be measured for smaller  angular separations
than some cutoff separation $\tmin$. For space-based observations,
this cutoff may be as small as one arc second. Ground-based
observations usually are limited by $\tmin \sim 5$ arcsec.

This inevitable cutoff at small angular scales leads to a biased
estimate of the aperture-mass statistics. We introduce the quantity
$\langle{M_{{\rm ap}, \perp}^2(\theta, \tmin)}\rangle$, which is defined as
in (\ref{Map-xi-EB}), but with the lower integration limit replaced by
$\tmin$. Even in the absence of a B-mode in the convergence field, a
$\tmin>0$ will create a spurious B-mode signal. For small angular
scales $\vt$, the first term in square brackets in (\ref{Map-xi-EB})
is approximately a positive constant, since both $\xi_+$ and
$T_+$ are integrals over a positive function multiplied by $\J_0$.
The second term tends to zero because of $\J_4$ in both $\xi_-$ and
$T_-$. This will result in a negative B-mode, $\langle
M_\perp^2(\theta, \tmin) \rangle < 0$, and an underestimation of the
E-mode, $\langle \Mapsq(\theta, \tmin) \rangle < \langle
\Mapsq(\theta) \rangle$. In fact, since the $T_-$-term is small for
small scales, the approximate relation $\langle M_{\rm ap}^2(\theta,
\tmin) \rangle - \langle M_\perp^2(\theta, \tmin) \rangle \approx \langle
M_{\rm ap}^2(\theta) \rangle$ holds for $\tmin/\theta \ll 1$.

Note that the same effects would arise if $\langle M_{\rm ap, \perp}^2
\rangle$ were obtained by putting apertures on the data field. In this
case, the estimator of the aperture-mass dispersion involves
summation over all galaxy pairs in an aperture. Due to the cutoff,
this sum would lack close pairs and the E-mode would be
underestimated.

\begin{figure}[!tb]
  
  \resizebox{\hsize}{!}{
    \includegraphics{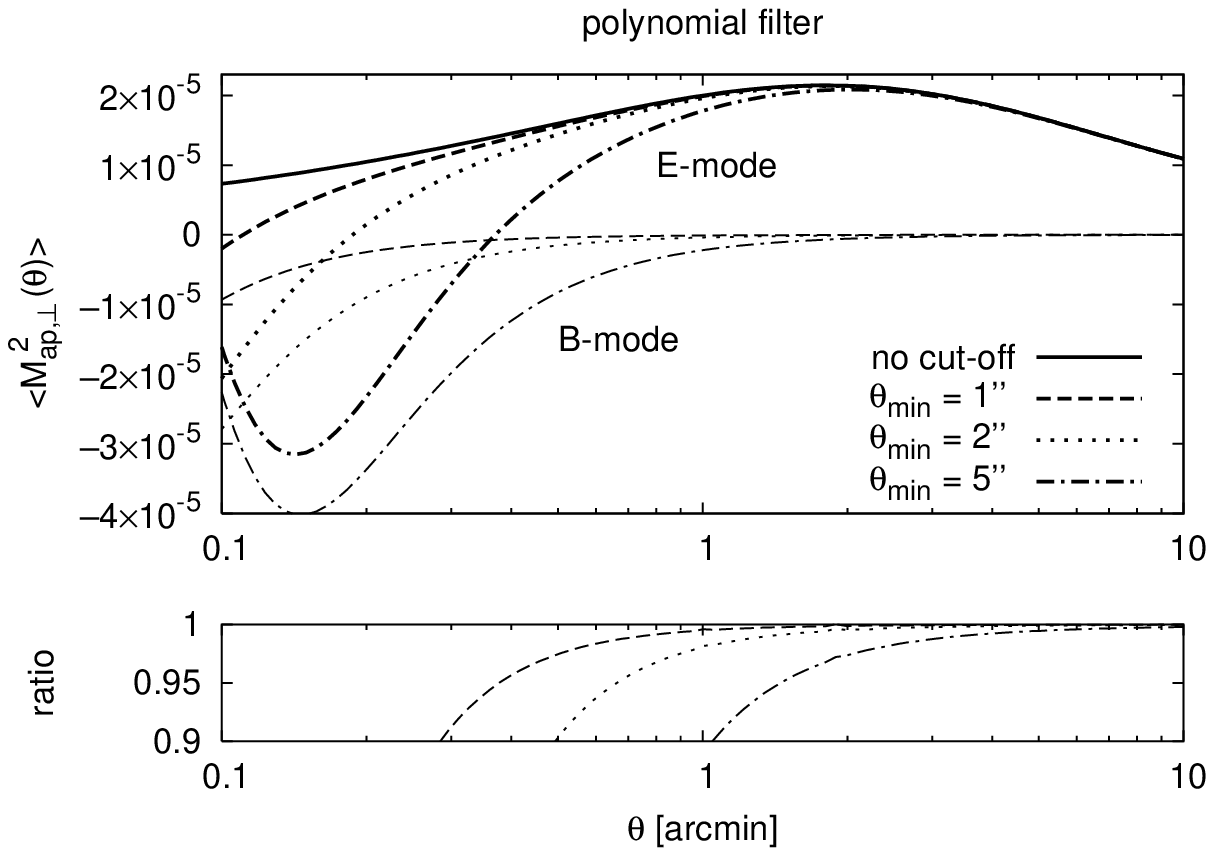}
  }

  \smallskip

  \resizebox{\hsize}{!}{
    \includegraphics{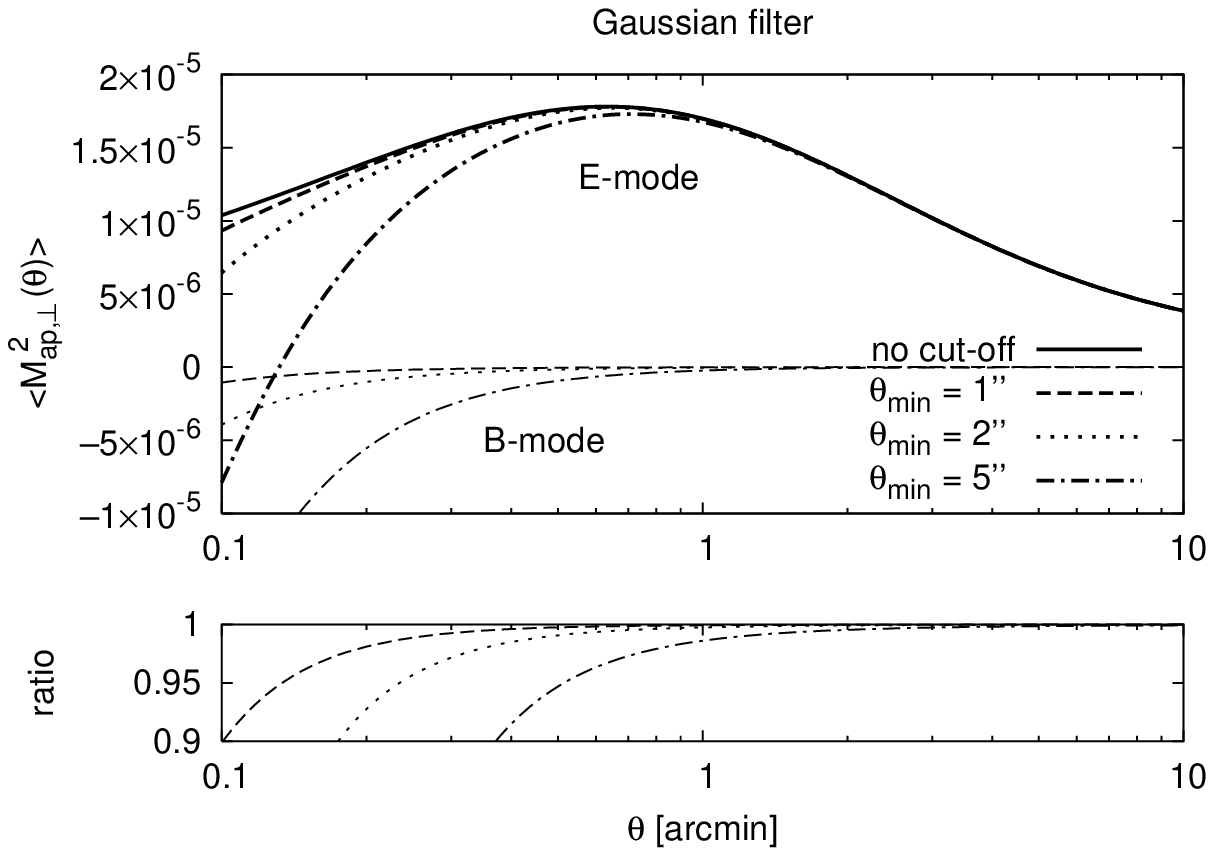}
  }

  \caption{The E-mode $\langle M_{\rm ap}^2(\theta) \rangle$ and the leakage
    from the E-mode $\langle M_{\rm ap}^2(\theta, \tmin) \rangle$ into
    the B-mode $\langle M_{\perp}^2(\theta, \tmin) \rangle$ due to the
    small-scale cutoff $\tmin$. The ratio $\langle
    M_{\rm ap}^2(\theta, \tmin) \rangle/\langle M_{\rm ap}^2(\theta)
    \rangle$ is plotted in the small panels.}
  \label{fig:mixing-p}
\end{figure}

\begin{figure}[!tb]
  
  \resizebox{\hsize}{!}{
    \includegraphics{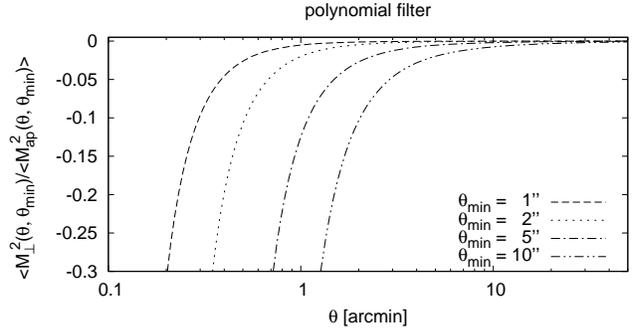}
  }
  
  \caption{The B- to E-mode ratio for various small-scale cutoffs
    $\theta_{\rm min}$.}
  \label{fig:MxMap}
\end{figure}

\begin{figure}[!tb]
  
    \begin{center}
      \resizebox{0.8\hsize}{!}{
        \includegraphics{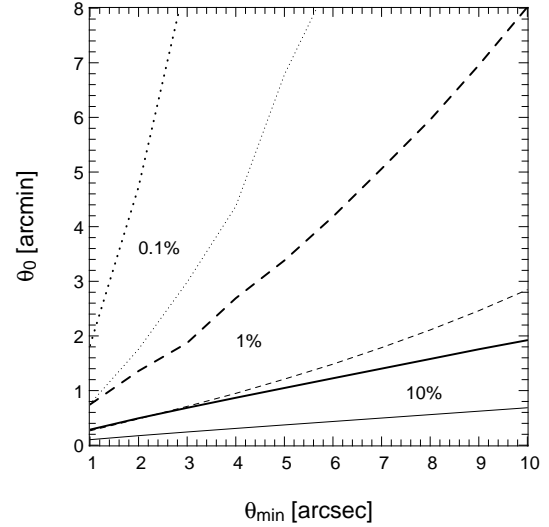}
      }
    \end{center}

  \caption{On scales $\theta$ larger than $\theta_0$, the ratio $\langle
    \Mapsq(\theta, \tmin) \rangle/\langle \Mapsq(\theta) \rangle$ is
    lower than $1 - p$, with lines corresponding to $p=10\%, 1\%,$ and
    $0.1$\% plotted as a function of the cutoff scale $\tmin$.
    Thick curves correspond to the polynomial filter (\ref{UQ-poly}),
    thin lines to the Gaussian functions (\ref{UQ-Gauss}).}
  \label{fig:theta0}
\end{figure}

Figures \ref{fig:mixing-p} and \ref{fig:MxMap} show the influence of
the cutoff $\tmin$ due to the absence of close pairs on the E- and
B-modes; see Sect.~\ref{sec:cosm} for the dependence on the
cosmological model and source redshift distribution. In
Fig.~\ref{fig:theta0} we quantify the suppression of the true E-mode.
Only for aperture radii $\theta$ larger than the threshold $\theta_0$
is the deviation smaller than indicated by the curves. For the
polynomial filter, one roughly finds $\theta_0 = 12 \tmin$ for 10\%
accuracy. For example, for ground-based observations with $\tmin = 5$
arc seconds one gets deviations of more than 10\% for $\theta \lsim
1\arcmin$. If a 1\%-precision is aspired scales smaller than
$3\farcm7$ have to be discarded.

The Gaussian filter is less affected by the cutoff than the
polynomial one, since it is broader and samples the 2PCF on larger
angular scales for a given aperture radius. However, this advantage
here turns into a disadvantage on large angular scales. There, the E-
and B-modes cannot be determined reliably due to the field boundary and
the infinite support of the Gaussian filter functions.

The definition of $T_+$ (\ref{T_pm}) shows that the mixing of the E-
and B-modes due to a small-scale cutoff is inevitable. Regardless of
the choice of the filter function, $U_\theta$, $T_+(x)$ tends to a
positive value for $x\rightarrow0$. 
The more rapidly $T_+$ falls off, the smaller the bias and thus the
shallower $\hat U$ and the broader $U_\theta$ is.

\section{E- and B-mode shear-correlation function}
\label{sec:xiEB}

The E- and B-mode correlation functions $\xi_{\rm E, B +}$ and
$\xi_{\rm E, B -}$ were defined in S02 and C02. They can be obtained
from the 2PCF in the following way,
\begin{eqnarray}
  \xi_{\rm E \pm}(\theta) & = & \frac 1 2 \left[ \xi_+(\theta) +
  \xi_-(\theta) + \xi_{\rm U, L}(\theta) \right]; \nonumber \\
  \xi_{\rm B \pm}(\theta) & = & \frac 1 2 \left[ \xi_+(\theta) -
  \xi_-(\theta) \mp \xi_{\rm U, L}(\theta) \right],
  \label{xiEB}
\end{eqnarray}
where the integral functions $\xi_{\rm U}$ and $\xi_{\rm L}$
correspond to the '$+$'- and '$-$'-cases, respectively, and are given by
\begin{eqnarray}
  \xi_{\rm U}(\theta) & = & \int_\theta^\infty \frac{\dd \vt}{\vt}
  \xi_-(\vt) \left( 4 - 12 \frac{\theta^2}{\vt^2} \right);
  \nonumber \\
  \xi_{\rm L}(\theta) & = & \int_0^\theta \frac{\dd \vt \vt}{\theta^2}
  \xi_+(\vt) \left( 4 - 12 \frac{\vt^2}{\theta^2} \right).
  \label{xiUL}
\end{eqnarray}

Thus, in order to separate the E- from the B-modes one needs to know
either $\xi_-$ up to very large or $\xi_+$ down to very small angular
scales. By an unfortunate conspiracy (or maybe not) $\xi_-$ falls off
rather slowly towards large scales, and $\xi_+$ is dominant for
$\vt\rightarrow 0$. Since a small-scale cutoff $\tmin$ due to close projected
galaxy pairs, as well as a maximum scale $\tmax$ due to the finite
observed field are unavoidable, an exact E-/B-mode separation using the
above-defined correlation functions is not possible.

Similar to the case of the aperture-mass dispersion
(Sect.~\ref{sec:map}), we define the functions $\xi_{\rm E,B+}(\theta,
\tmax)$ and $\xi_{\rm E,B+}(\theta, \tmin)$. In the first case, the
infinite integral for $\xi_{\rm U}$ (\ref{xiUL}) is truncated at
$\tmax$; in the second case, the lower integration limit for $\xi_{\rm
  L}$ is replaced by $\tmin$.

\begin{figure}[!tb]
  
  \resizebox{\hsize}{!}{
    \includegraphics{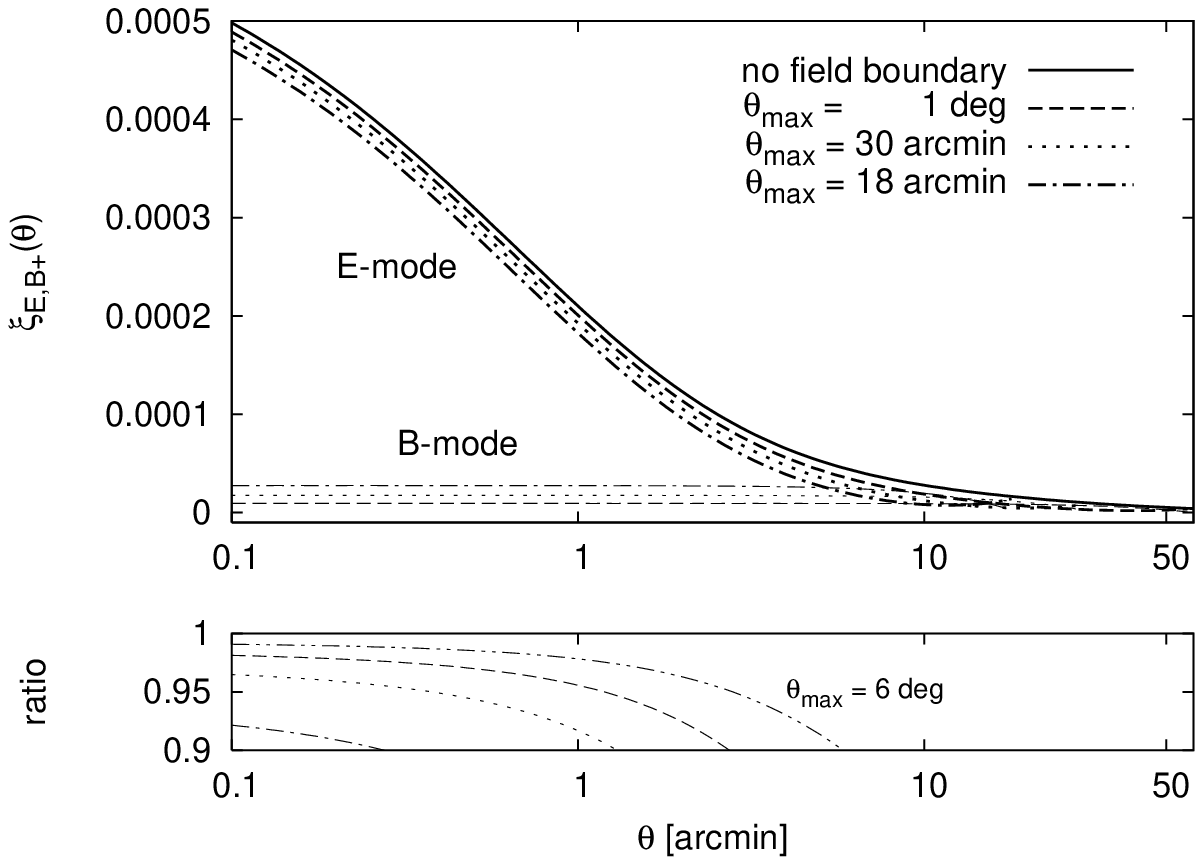}
  }

  \resizebox{\hsize}{!}{
    \includegraphics{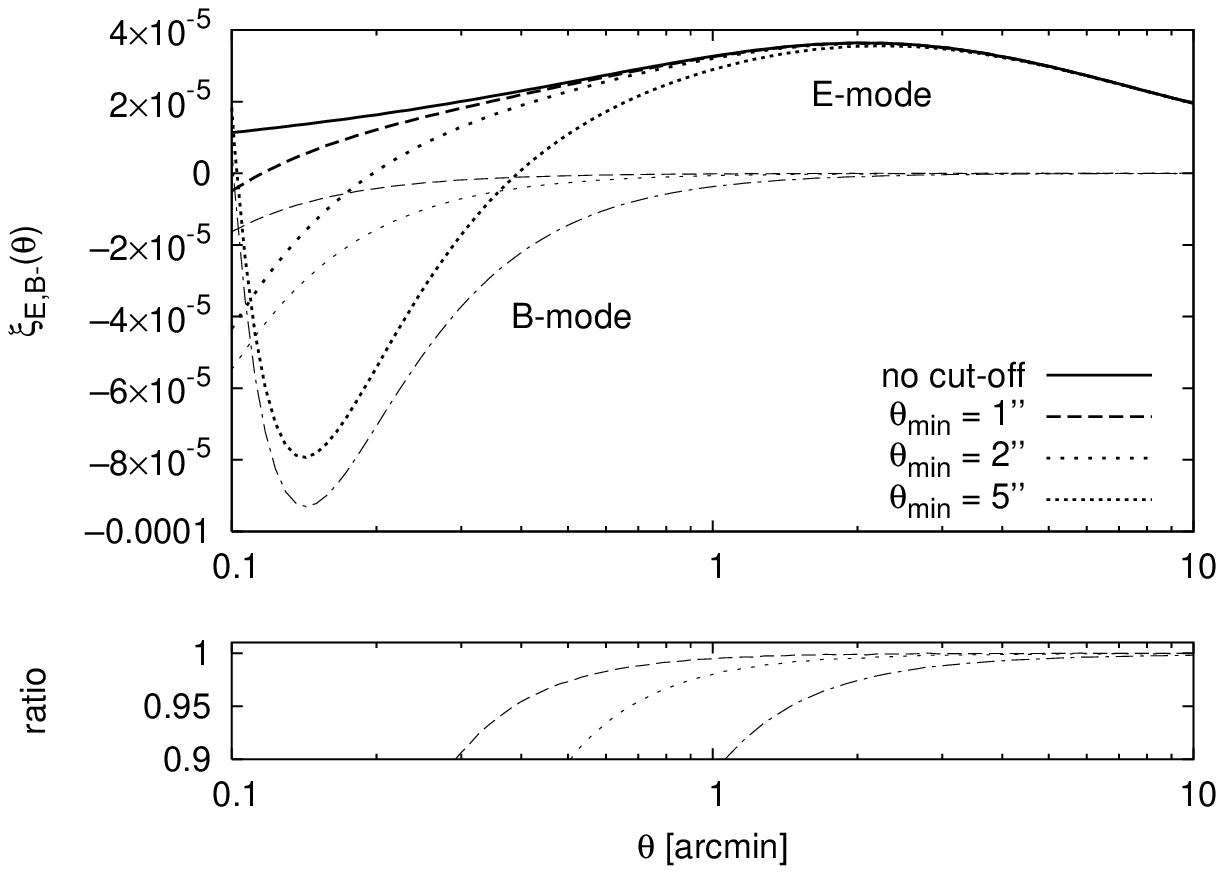}
  }
  
  \caption{The E-/B-mode correlation functions. \emph{Top panels:} Mixing
    between $\xi_{\rm E+}$ and $\xi_{\rm B+}$ due to a finite field.
    The curves correspond to an infinite field and to fields with
    maximal scales of $\tmax=$ 1 deg, 30, and 18 arc minutes as indicated
    in the panel. (The E- and B-mode for $\tmax=6$ deg is not shown in the
    upper panel.) \emph{Bottom panels:} Mixing
    of $\xi_{\rm E,B-}$ due to a small-scale cutoff $\tmin$. In the
    small panels, the ratio between the corrupted and the true E-mode
    is shown.}
  \label{fig:xiEB}
\end{figure}

In Fig.~\ref{fig:xiEB}, the mixing of the E- and B-mode correlation
functions is displayed. The large-scale cutoff leads to an
underestimation of the true E-mode on all scales, even down to arc
seconds. At the same time, a spurious B-mode $\xi_{\rm B+}$ is
introduced. Even for a maximum angular separation of 1.5 degrees, the
E-mode is off by more than one percent on all relevant scales. The
`minus'-E-mode $\xi_{\rm E-}$ suffers badly from the close pair
cutoff on scales as large as several arc minutes.

\begin{figure}[!tb]
  
  \begin{center}
    \resizebox{0.8\hsize}{!}{
      \includegraphics{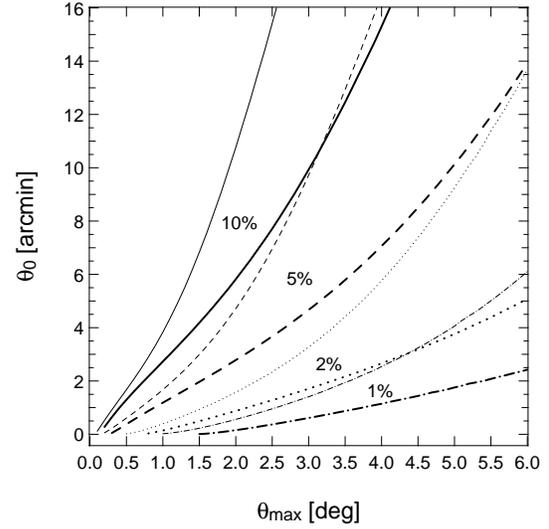}
    }
  \end{center}
  
  \caption{On scales $\theta$ smaller than $\theta_0$, the ratio
    $\xi_{\rm E+}(\theta, \tmax)/\xi_{\rm E+}(\theta)$ is lower than
    $1-p$, where lines corresponding to $p=10, 5, 2,$ and 1\% are
    plotted as a function of the maximum separation $\tmax$. Thick
    curves correspond to a mean source redshift of $\bar z = 1$, thin
    lines to $\bar z = 1.5$.}
  \label{fig:theta0-xiEp}
\end{figure}

\begin{figure}[!tb]
  
  \begin{center}
    \resizebox{0.8\hsize}{!}{
      \includegraphics{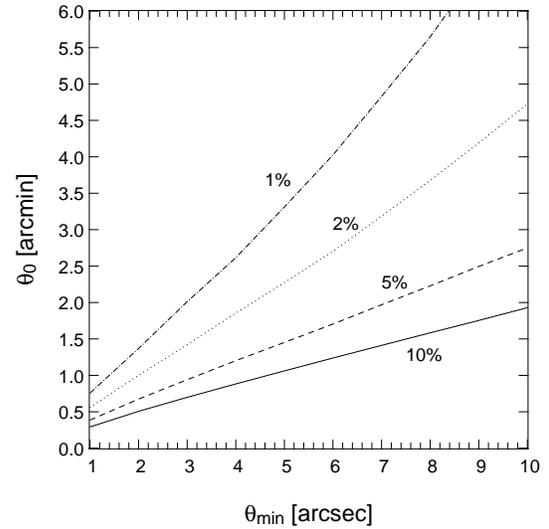}
    }
  \end{center}
  
  \caption{On scales $\theta$ larger than $\theta_0$ the ratio
    $\xi_{\rm E-}(\theta, \tmax)/\xi_{\rm E-}(\theta)$ is smaller than
    $1-p$, where lines corresponding to $p=10, 5, 2$ and 1\% are
    plotted as a function of $\tmin$.}
  \label{fig:theta0-xiEm}
\end{figure}

Figures \ref{fig:theta0-xiEp} and \ref{fig:theta0-xiEm} show the
deviation of the corrupted E-mode $\xi_{\rm E+}(\theta, \tmax)$ and
$\xi_{\rm E-}(\theta, \tmin)$ from the true E-mode $\xi_{\rm
  E+}(\theta)$ and $\xi_{\rm E -}(\theta)$, respectively.

\section{Dependence on cosmology and source galaxy redshift}
\label{sec:cosm}

The amplitude and the shape of the shear statistics considered here
depend, of course, on the underlying cosmology and the resdhift
distribution of source galaxies. However, the ratio of the observed
E-mode in the presence of the small-scale cutoff $\tmin$ to the true
E-mode ($\langle M_{\rm ap}^2 \rangle$, $\xi_{\rm E, B-}$) is very
insentive to the convergence power spectrum. In the case of $\xi_{\rm
  E, B+}$, the difference between the E-mode corresponding to a
maximal scale $\tmax$ and the true E-mode depends only little on
the underlying model. However, the amplitude of $\xi_{\rm E+}$ depends
quite strongly on the amplitude of the power spectrum, which is
governed in particular by $\Omegam, \sigma_8$, and the source redshift
distribution. As a consequence, the ratio of the observed to true
E-mode depends on those parameters. For a deeper survey, this ratio
will be lower than for a more shallow one. For all plots in this
paper, if not indicated otherwise, a mean redshift of $\bar z = 1.0$
is assumed. The cosmology is $\Lambda$CDM with $\Omegam = 0.3$ and
$\sigma_8 = 0.85$.

\section{Summary and conclusions}
\label{sec:summary}

We have quantified the mixing between the E- and the B-modes which
arises due to the lack of information about the shear correlation on
both very small and very large scales. The former limit $\tmin$ is due
to close projected pairs, the shape of which cannot be determined
reliably. The latter cutoff $\tmax$ is related to the finite size of
the observed galaxy fields. Apart from that, a fundamental limit exists
since only a $4\pi$-survey can in principle yield an unambiguous E-
and B-mode decomposition which, however, is unrealistic due to the
occlusion of parts of the sky by the Milky Way
\citep{2003NewAR..47..987B}. We used the aperture-mass dispersion
$\langle M_{\rm ap, \perp}^2 \rangle$ and the shear-correlation
functions $\xi_{\rm E,B \pm}$ as measures of the E- and B-modes,
respectively.

Even if the cutoff due to close galaxy pairs is as small as
$5\arcsec$, which is feasible for ground-based observations, the E-mode
is underestimated by 10\% (1\%) on scales below $1\arcmin$
($3\farcm4$), using $\langle M_{\rm ap}^2 \rangle$ with the polynomial
filter function. Moreover, a negative B-mode signal appears. Since
even with space-based measurements, a cutoff of about
$\tmin=1\arcsec$ exists due to blending galaxy images, it is
fundamentally not possible to know whether there is a B-mode present
on scales smaller than one or two arc minutes with very high
(sub-percent) precision. This cutoff does not
introduce a B-mode into the 2PCF, however, it prevents the clear
detection of any B-mode still present in the data that may corrupt the
2PCF on much larger scales than $\tmin$.
  
The E-/B-mode correlation function $\xi_{\rm E, B-}$ behaves in a
similar way than $\langle M_{\rm ap}^2 \rangle$. The `$+$'-version,
$\xi_{\rm E, B+}$, on the other hand is affected by a large-scale
cutoff $\tmax$ that causes a leakage from the E- to the B-mode on
all angular scales. For fields smaller than 1.5 degrees, the precision
of the E-mode correlation function is never better than 1\%.

In order to obtain knowledge about the B-mode in the data, one has to
combine $\xi_{\rm E, B+}$ and $\xi_{\rm E, B -}$. For ground-based
data ($\tmin = 5\arcsec$), $\xi_{\rm E, B -}$ allows one to
constrain the B-mode at the percent level on scales larger than
$3\farcm3$. To get this precision on larger scales using $\xi_{\rm E,
  B +}$, the measurement of the shear correlation up to about $\tmax =
7$ degrees is necessary. From space-based observations with $\tmin =
1\arcsec$, $\tmax = 3.4$ deg ($\tmax = 2.4$ deg) is needed for a
mean redshift of $\bar z = 1.0$ ($\bar z = 1.5$), respectively.

Up to now, various strategies have been employed to estimate the E- and
B-mode from shear data. \citet{2005MNRAS.359.1277M} calculate the
E-/B-mode correlation functions by extrapolating $\xi_-$ beyond
measured scales, using a theoretical model for $P_\kappa$. However,
this method explicitely makes the assumption that no B-mode is present
on larger scales than probed by the survey. Choosing a wrong model can
change the amplitude of $\xi_{\rm E}$ on the measured angular range,
which might bias the cosmological interpretation of the shear
correlation. In particular, $\Omegam$ and $\sigma_8$ might be biased,
which, although within the systematic errors for current measurements
\citep[see Fig.~10 of][]{2005MNRAS.361..160H}, will be problematic for
future surveys.

Recently, \citet{vWMH05} use the aperture-mass dispersion to calibrate the
E-mode shear-correlation function. Since their measured
B-mode is consistent with zero on large scales ($10\arcmin$ --
$50\arcmin$), they shift $\xi_{\rm E, B}$ such that the B-mode
vanishes on corresponding angular scales ($2\arcmin$ -- $10\arcmin$).
It has to be noted that a B-mode on very large scales does not have any
influence on $\langle M_{\rm ap}^2 \rangle$ on smaller scales, but it
will corrupt the correlation function. Moreover, there is no clear
correspondence between the angular scales probed by the correlation
function and the aperture-mass dispersion.

Alternatively, the aperture-mass statistics can be measured and fitted
with a cosmological model \citep[e.g.][]{J03}. In that case, either
scales below $\theta_0$ have to be omitted from the analysis, or the
theoretical prediction of $\langle M_{\rm ap}^2 \rangle$ has to be
obtained using eq.~(\ref{Map-xi-EB}) with the cutoff $\tmin$
specified by the observations. Note, however, that in this case
the B-mode signal should also be fitted with the predicted, non-zero
$\langle M_\perp(\theta, \tmin) \rangle$ if no lensing information is
to be lost.

At present, the prediction on the shear signal at angular separations
on arc-minute scales and below is very difficult and inaccurate.
These small scales should not be disregarded since the shear signal on
these scales is very high and contains unique information about the
non-linear and non-Gaussian features of the large-scale structure.  In
the future, when our non-linear models of structure formation improve
and baryonic effects are taken into account, shear measurements on
small scales will be of great use.  Since the shear signal on small
scales is particularly susceptible to contamination by intrinsic
galaxy alignment, a clear separation between the E- and B-mode on
these scales is mandatory. Only then can one make certain that the
measurement is free of systematic errors and has the necessary quality
to be used for precision cosmology.

\section*{Acknowledgments}

We thank Marco Hetterscheidt, Tim Schrabback-Krahe, and Richard Massey
for fruitful discussions, and the anonymous referee for useful
comments.

\bibliographystyle{aa} \bibliography{astro}

\end{document}